%% file: taylor.tex
\documentclass[12pt]{rec06}
\usepackage[latin1]{inputenc}
\usepackage[english]{babel}
\usepackage{url}
\usepackage{fullpage,supertabular,alltt,latexsym,pvs}
\begin{document}
\begin{article}
\begin{opening}

\title{A library of Taylor models for PVS automatic proof checker}
%
%
\author{{\bf Francisco \surname{Cháves}}\footnote{This material is  based  on work supported by the
    \emph{Mathlogaps} (Mathematical Logic and Applications) project, an
    Early Stage Research Training grant of the European Union.
}
  and   {\bf Marc \surname{Daumas}}\footnote{This  work   has  been  partially
    supported by   PICS  2533 from   the French  National   Center for
    Scientific Research (CNRS).}}
\runningauthor{Francisco Cháves and Marc Daumas}
\runningtitle{A library of Taylor models for PVS automatic proof checker}

\institute{\bigskip Laboratoire de l'Informatique du Parallélisme\\%
           UMR 5668 CNRS--ENS de Lyon--INRIA\\%
           email: Francisco.Jose.Chaves.Alonso@ENS-Lyon.Fr}
\institute{Laboratoire d'Informatique, de Robotique et de Microélectronique de Montpellier\\%
           UMR 5506 CNRS--UM2\\%
email: Marc.Daumas@LIRMM.Fr}
\institute{Visiting Laboratoire de Physique Appliquée et d'Automatique\\%
           EA 3679 UPVD\\%
           email: Marc.Daumas@Univ-Perp.Fr}
\date{}

\begin{abstract}
  We present in this paper a library  to compute with Taylor models, a
  technique extending interval arithmetic  to reduce decorrelation and
  to solve differential equations. Numerical software usually produces
  only numerical results.  Our library   can be used to produce   both
  results and proofs.   As seen during  the development of  Fermat's
  last theorem reported  by \citeauthor{Acz96} \shortcite{Acz96},  providing a proof
  is  not sufficient.   Our  library provides  a  proof that  has been
  thoroughly scrutinized by a trustworthy and tireless assistant.  PVS
  is an automatic proof assistant that has  been fairly developed and
  used and that has no internal connection with interval arithmetic or
  Taylor models.   We  built our library  so that  PVS validates  each
  result as  it is produced.  As producing  and validating a proof, is
  and  will certainly  remain a   bigger  task than  just producing  a
  numerical   result  our library  will  never   be  a  replacement to
  imperative implementations of Taylor  models such as Cosy  Infinity. 
  Our library should  mainly be used to  validate small to medium size
  results that are involved in safety or life critical applications.
\end{abstract}

\keywords{PVS, program  verification, interval   arithmetic,
Taylor models.}
\end{opening}

\section{Introduction}
\label{sec:intro}

Taylor models, see for example  \cite{MakBer03} and references herein,
have  recently emerged  as  a  nice   and  convenient way to    reduce
decorrelation           in          interval                arithmetic
\cite{Moo66,Neu90,JauKieDidWal01}.   Taylor     models  are even  more
attractive   when one solves initial value   problems  ODEs as they
provide a validated built-in integration operator.


Yet, it is now beyond doubt that  programs and libraries contain bugs,
no matter how precisely they  have  been specified and how  thoroughly
they  have been tested  \cite{RusHen91,Ros05}.   As a consequence, the
highest    Common     Criteria   Evaluation      Assurance      Level,
EAL~7\footnote{\url{http://niap.nist.gov/cc-scheme/}.}, has only  been
awarded so   far to products that  provide  validation using  a formal
tool, specifically an automatic proof checker in first or higher order
logic.


We    present  here    our library    of      Taylor  models  in   PVS
\cite{OwrRusSha92}. Working with an automatic proof checker, we had to
manage two   tasks.  The first task  was  to  create a  data  type and
operations  on this new  type to  allow users  to define and  evaluate
expressions using   Taylor  models.  The second    task was to provide
proofs that each   operator is correct  and a  strategy to recursively
analyze    compound expressions.   Both  tasks  rely   on the recently
published library  on interval arithmetic  for PVS \cite{DauMelMun05}. 
As many   mathematical developments are not yet   available in PVS, we
also had to develop an extended library on  polynomials and prove a few theorems
of analysis and algebra.


Our library on Taylor models  can be used   to  derive quickly  more or  less accurate
bounds. For example, users of formal  tools have to provide proofs
that    radicals are non negative   for all  expressions using
square roots.  Some proofs  use intricate analysis but most of
them are very simple and interval arithmetic or low degree evaluations
with  Taylor models can produce   appropriate proofs. Our library  can
also be used to expertly derive computer validated proofs of difficult
results through an expert use of Taylor models.


The library will be available freely on the Internet  as soon as it is
stable.  Side developments are  integrated  as they are  produced  to
NASA                             Langley                           PVS
libraries\footnote{\url{http://shemesh.larc.nasa.gov/fm/ftp/larc/PVS-library/pvslib.html}.}.
Meanwhile, all files can be retrieved from the author's website.
\begin{center}
\url{http://perso.ens-lyon.fr/francisco.jose.chaves.alonso/pvs-files/}
\end{center}

\subsection{Working with an automatic proof checker}
\label{sub/pvs}

Software is used extensively for a wide array of tasks. Some pieces of
software should never fail.  The ones used by transportation means 
(planes, buses, cars\ldots),  for   medical care  (controlling   pumps,
monitors,   prescriptions\ldots) or  in  the  army   (parts of  weapons,
alarms\ldots) belong to the  fast lengthening list  of life  or safety
critical applications.   A   mindless  modification of  one  parameter
reportedly   caused human  losses  in  the  \emph{Instituto Oncologico
  Nacional} on Panama  where eight people  died and twenty others were
hurt     \cite{GagMcC04}.   Many lethal   and  costly  failures
\cite{USA.92,Lio.96}  show  beyond reasonable doubts that traditional
software verification is not sufficient to guarantee correct behavior.

PVS\footnote{\url{http://pvs.csl.sri.com/}.}   (Prototype Verification
System) by  \citeauthor{OwrRusSha92} \shortcite{OwrRusSha92,OwrShaRusStr01b,OwrShaRusStr01a} is    one
environment for the development  and analysis of formal specifications
that allows the elaboration of  theories and proofs. The system 
deals with  theories where  users develop definitions, axioms
and theorems.  To verify  that 
theorems are correct, PVS uses  a typed  higher order logic
language where new  types  are defined  from a list  of basic  types
including booleans, natural numbers, integers\ldots 
The   type system   allows the  definition  of
functions, registers, tuples and abstract data types.

PVS uses  \emph{predicate subtype}s, 
subtypes where all objects  satisfy a given
predicate.  For  example $\{ x:\text{real}| x  \neq 0\}$ is the set of
non--zero reals.   Subtype  predicates  are used  for operations  that
aren't defined for all possible inputs.  This restriction is therefore 
visible in
the signature  of  the  operation.   For example  the  division  is an
operation of real numbers  such that the  type  of the denominator  is
a  real number  different from zero.   As a result,  all
functions of PVS   are total in  the sense  that  the  domain  and the
signature must exclude explicitly any  input where a function could
not be defined.

As  predicates  used  by the   system to define  types are
arbitrary,  type verification  is    undecidable and  it   usually
generates proofs  obligations named \emph{type correctness conditions}
(TCCs). Users have to provide proofs  of generated TCCs with the
help of  PVS.  

In   PVS the  $\lambda$  operator  defines  anonymous functions.   
Expression  $\lambda x.e$ is a   function that  has parameter $x$  and
returns expression $e$.  For example, the function that returns $0$
for any value of its single parameter could be defined as $\lambda
x.0$ and  identity function that  returns the same element that is
given as parameter is $\lambda x.x$.  Function {\tt $\lambda$ k :
  nat. if k  = 0 then  1 else 0} is the  sequence that for  input $0$,
returns $1$, and returns $0$ for any other input.

Nowadays, systems such as PVS are  fully able to certify that programs
are corrects \cite{Ros05} but  programmers scarcely use them.  Providing
a formal proof of correct behavior is  a difficult task, it requires a
specific training   and user interfaces   of proof  assistants are  of
little help for all the work that is  not done automatically.  Hope is
that as more and more work is done automatically, users will need only
limited    interactions with automatic   proof  checkers down to the
point   where no  interaction  is required   at all.   This trend  was
recently coined as \emph{invisible formal methods} \cite{TiwShaRus03}.

\subsection{A few words about interval arithmetic}

In  interval  arithmetic scalar variables  $x$  are  replaced by pairs
$(a,b)$ with the semantic that $x$ lies in the interval $[a,b]$. Later
on, we compute  bounds rather than   values. We use operators commonly
found in programming languages  such as addition, subtraction,
multiplication and so on \cite{JauKieDidWal01}.
$$
\begin{array}{r c l}
~ [a,b] +     [a',b'] &=& [ a +     a', b +     b' ] \\
~ [a,b] -     [a',b'] &=& [ a -     b', b -     a' ] \\
~    c  \cdot [a ,b ] &=& [ c \cdot a , c \cdot b  ] \qquad c \geq 0 \\
~ [a,b] \cdot [a',b'] &=& [ \min \{ a a', a b', b a', b b'\}, \max \{ a a', a b', b a', b b'\} ]
\end{array}
$$

Working with automatic proof checkers, we convert operations into properties \cite{DauMelMun05}.

\begin{center}
For all $x \in [a,b]$, $y \in [a',b']$ and $c \in \mathbb{R}$
$
\left\{
\begin{array}{r c l}
  x + y     & \in & [a,b] + [a',b']     \\
  x - y     & \in & [a,b] - [a',b']     \\
  c \cdot x & \in &    c  \cdot [a,b]   \\
  x \cdot y & \in & [a,b] \cdot [a',b']
\end{array}
\right.
$
\end{center}

Decorrelation is a problem intrinsic to interval arithmetic.
There is decorrelation on interval evaluation of any expression
where one or more  variables appear more than once.
For example, the most simple scalar expression
\[
x - x
\]
where $x \in [0,1]$, is replaced in interval arithmetic by
\[
   [0,1] - [0,1] = [-1,1].
\]

Everyone agrees that $x - x$ lies in the interval $[0, 0]$ but interval arithmetic
produces the correct but very poor $[-1,1]$ interval.
Decorrelation and other problems lead  interval arithmetic
to overestimate the domain of
results. Techniques are used intensively to produce constrained  results.

One  of such  techniques  is based on   Taylor's theorem with Lagrange
remainder where $f$ is  $n$ times continuously derivable between $x_0$
and $x$, $f$  is $n+1$ times  derivable strictly between $x_0$ and $x$
and $0 < \theta < 1$.
$$
\begin{array}{r c l}
\displaystyle  f(x) & = & f(x_0)+(x - x_0)f'(x_0) + \frac{(x - x_0)^2}{2!}f''(x_0)     \\[6pt]
\displaystyle       & + & \cdots ~ + \frac{(x - x_0)^n}{n!}f^{(n)}(x_0)                  \\[6pt]
\displaystyle       & + & \frac{(x - x_0)^{n+1}}{(n+1)!}f^{(n+1)}(x_0+(x - x_0)\theta )
\end{array}
$$
Adapting Taylor's theorem to interval arithmetic, we obtain the
formula below \cite{DauMelMun05} for $x$ and $x_0$ in $I$.
$$
\begin{array}{r c l}
\displaystyle  f(x) & \in & f(x_0)+(I - x_0)f'(x_0) + \frac{(I - x_0)^2}{2!}f''(x_0) \\[6pt]
\displaystyle       & +   & \cdots ~ + \frac{(I - x_0)^n}{n!}f^{(n)}(x_0)            \\[6pt]
\displaystyle       & +   & \frac{(I - x_0)^{n+1}}{(n+1)!}f^{(n+1)}(I )
\end{array}
$$

Using Taylor's theorem was appropriate  in \cite{DauMelMun05} but it has many drawbacks:
\begin{itemize}
\item It is difficult to hide the use  of Taylor's theorem in order to
  provide {\em  invisible formal methods}.   This is due  to the large
  number  of quantities involved in  instantiating  the theorem in its
  generic  form.  Progress  has  been   achieved  by  Muñoz after   the
  publication of \citeauthor{DauMelMun05}.
\item To use Taylor's theorem,  one has to  express the derivatives of
  function $f$.
\item For   large  expressions,  $f$ alone  might  be too large   to be
  expressed in PVS.
\end{itemize}

Taylor models presented in the rest of this text overcome all the previous
drawbacks to the price of a less accurate approximation.
We have developed a set operations for PVS that includes  addition,
negation,  scalar     multiplication, multiplication,  reciprocal  and
exponential. 
We  present our   developments in PVS, first quickly  on polynomial functions and then on  Taylor models.
We finish  with  concluding remarks 
and a few toy examples.

\section{Implementing polynomials in PVS}


For the implementation of polynomials we considered a finite list of
monomial functions, a finite sequence of coefficients and an infinite
power series with finite support.  Finite lists or sequences usually
imply the construction of a new inductive type {\em à la} Coq\footnote{See for example
  \url{http://www.lfcia.org/staff/freire/phd-gilberto/gilberto_phd_html/}.} \cite{BerCas04}.
We implemented polynomials as power series with finite support. This
scheme is appropriate for a proof system like PVS and is compatible
with NASA series
libraries\footnote{http://shemesh.larc.nasa.gov/fm/ftp/larc/PVS-library/pvslib.html.}.

Working with sequences of coefficients rather than monomial functions
means that we need the {\tt powerseries} function to evaluate
polynomial $P$ on input $x$.  It also means that some theorems can be
established on finite support series rather than polynomial functions.

%

\subsection{Finite support series}
\label{sub/fs}

\begin{figure}
  \begin{center}\fbox{\begin{minipage}{15cm}
  \input{finite_support}
  \end{minipage}}\end{center}
  \caption{Abridged and reordered theory on finite support series (see file {\tt finite\_support.pvs})}
  \label{pvs/fs}
\end{figure}  

Our implementation  of polynomials is outlined in Figure~\ref{pvs/fs}. 
It mostly
describes mathematical objects (definition, function, theorems...) with
common words except for the notions introduced in Section~\ref{sub/pvs}

We
define predicate {\tt finite\_support (a,N)} just after the
preamble. Addition of sequences was already defined and is imported
from previous work in the preamble. We had to define a product
operator and a composition operator. The first operator  applies to generic
series. The second operator requires that the first sequences $a$ returns  zero for
indices above input $d$.




In the second half of Figure~\ref{pvs/fs}  we proved that
negation, addition, multiplication by a scalar,
multiplication and composition 
 return finite support series provided (both) inputs are
finite support series. We also proved that Cauchy's product is meaningful
for finite support series. The meaning of composition can only be assessed in
regard to polynomial functions.

\subsection{Polynomial}

\begin{figure}
  \begin{center}\fbox{\begin{minipage}{15cm}
  \input{polynomials_ext}
  \end{minipage}}\end{center}
  \caption{Abridged extensions to the theory on polynomial (see file {\tt polynomials\_ext.pvs})}
  \label{pvs/pol}
\end{figure}

As we have mentioned  earlier, we use {\tt  polynomial (a, n)} function
to  create a power  series from finite  support  sequence $a$ based on
{\tt powerseries(a)(x)(N)}  function implemented in previous work.   Extended results on polynomial
functions are presented in Figure~\ref{pvs/pol} based on NASA libraries.
$$ 
  \text{polynomial}(a,n)(x) =  \sum_{k = 0}^n a_k \cdot x^k
$$

We proved in this file that Cauchy's multiplication applies to finite
support series as well as polynomial functions. We also proved that
the series obtained from composing two finite support series
as defined in Section~\ref{sub/fs} defines the same polynomial function
as the one that would be obtained by composing the polynomial
functions associated to the two initial series.

Technical results are also presented in this file to provide more
insights to our development.

\section{Taylor models}

\begin{figure}
  \begin{center}\fbox{\begin{minipage}{15cm}
  \input{taylor_model}
  \end{minipage}}\end{center}
  \caption{Abridged and reordered theory on Taylor models (see file {\tt taylor\_model.pvs})}
  \label{pvs/tm}
\end{figure}  

Taylor models \cite{MakBer03} are pairs $t = (P,I)$ where $P$ are
polynomial functions of fixed degree $N$ and $I$ are intervals. $N$ is
a constant that cannot be changed during the evaluation of
expressions. In PVS, pairs are defined using components between
{\tt (\#} and {\tt \#)}. Components can be addressed independently
using quotes {\tt `}, that are {\tt t`P} and {\tt t`I}.

Taylor model $t$  is a correct representation of  function $f$ if it 
satisfies the {\tt containment} predicate stated Figure~\ref{pvs/tm},
$$
\forall x \in J ~~~~~ f(x) - t'P(x) \in t'I
$$
where $J$ is usually $[-1,1]$.



Our first task was to define operations on Taylor models. Addition,
negation and multiplication by a scalar are straight forward and can
be read directly from Figure~\ref{pvs/tm}.
Naive multiplication of Taylor models creates polynomials of degree
$2N$. The high order terms of the polynomials must be truncated and are
accounted for in the interval part.

The {\tt inv} reciprocal operator uses the following equality where $r \in I$, $p(0) \neq 0$ and $p(x)$ has the same sign as $p(0)$.
\begin{equation}
\frac{1}{p(x) + r} = \frac{1}{p(0)} \cdot \frac{p(x)}{p(x) + r} \cdot \frac{1}{1 - \left(1 - \frac{p(x)}{p(0)}\right)} \\
\label{ieqn/rec} 
\end{equation}
We define  $q(x) = 1 -  \frac{p(x)}{p(0)}$  and we  expand the last
fraction of (\ref{eqn/rec})   using the   geometrical series    $\sum_{i=0}^N q^i$
truncated to keep only a polynomial of degree $N$.

Decorrelation forbids to evaluate the penultimate fraction of
(\ref{eqn/rec}) directly and we defined a new operator based on the
lower bound and the upper bound of $I/p(J)$ that returns directly
$$
\left[ \frac{1}{1+\frac{1}{lb'(I/p(J))}}, \frac{1}{1+\frac{1}{ub'(I/p(J))}} \right].
$$
This operator cannot be replaced by a direct implementation of 
$$
\frac{1}{1 + p(J) / I} ~~ \text{or} ~~ \frac{1}{1 + \frac{1}{I / p(J)}}
$$
because $I$ usually contains $0$ preventing anyone to use it as a divisor.


We also implemented the exponential of Taylor models using the
following equality where $r \in I$ and $\hat{e}^x$ is a rational
approximation of $e^x$.
$$
e^{p(x) + r} = \hat{e}^{p(0)} \cdot e^{p(x) - p(0)} \cdot \frac{e^{p(0)}}{\hat{e}^{p(0)}} \cdot e^r
$$
The polynomial part of the result is obtained by developing and
truncating the exponential series composed with $p(x) - p(0)$. The interval part is set
accordingly to account for all discarded quantities.
 



The five {\tt \_sharp} lemmas of the second part of Figure~\ref{pvs/tm},
show that the {\tt containment} predicate is preserved by our 
operators. It  means that we can deduce properties from evaluations of
expressions using Taylor models.









In addition to prove mathematical theories, PVS provides a \emph{ground evaluator}.
It is an experimental feature of PVS that enables the animation
of functional specifications. To evaluate them,
the ground evaluator extracts Common Lisp code
and then evaluates the code generated on PVS underlying
Common Lisp machine.

Uninterpreted PVS functions can be written in Common Lisp. PVS only trusts
Lisp codes generated automatically from PVS functional
specifications, then one can not introduce
inconsistencies in PVS. However, codes
are not type-checked by PVS and can break inadvertently.
 
PVSio\footnote{http://research.nianet.org/~munoz/PVSio}
 is a PVS package developed by Muñoz that extends the ground
evaluator with a predefined library including imperative
programming language features. PVSio  loads in  emacs interface using 
{\tt M-x load-prelude-library PVSio}  and then executes with {\tt M-x
pvsio}. 

\section{Toy example, concluding remarks and future work}

\begin{figure}
  \begin{center}\fbox{\begin{minipage}{15cm}
        \input{examples}
  \end{minipage}}\end{center}
  \caption{A toy example of Taylor models (see file {\tt example.pvs})}
  \label{pvs/ex}
\end{figure}  

Figure~\ref{pvs/ex} show how easily we can define expressions. 
PVSio is used to evaluate Taylor model expressions and 
Figure~\ref{pvs/trace} shows the  polynomial and
interval parts of the Taylor model of degree 5 of
$$
ch \left( 2 \cdot \frac{x}{1000} \right) \cdot sh \left( 3 \cdot \frac{x}{1000} \right) =
3 \cdot \frac{x}{1000} + \frac{21}{2} \cdot \left(\frac{x}{1000}\right)^3 + \frac{521}{40} \cdot \left(\frac{x}{1000}\right)^5 + r
$$
with 
$$
r \in 5150892483 \cdot 10^{-28} \cdot [-1, 1]
$$

Coefficients are obtained from  expressions
{\tt example1`P(0)}, {\tt P(1)} down to  {\tt
P(5)}. The interval part is {\tt example1`I}.

\begin{figure}
  \begin{center}\fbox{\begin{minipage}{15cm}
        {\tt \input{examplesTrace}}
  \end{minipage}}\end{center}
  \caption{Trace of our toy example of Taylor models}
  \label{pvs/trace}
\end{figure}

To conclude, we would like to  say that they have three goals
in presenting this report:
\begin{itemize}
\item {\bf  Present an accurate  report of the work involved including
    the training  of   a PhD student to PVS}.    Though  this development  is
  significant,  PVS validated  projects    can  be achieved  in    a
  reasonable time-frame provided appropriate tutoring is available.
\item {\bf Provide a simple tutorial to our library on Taylor models}.
  Readers should be able to start validating their  own results as soon as they
  have finished reading this paper.
\item  {\bf Offer a first  easy step to   the usage of automatic proof
    checkers}. It is always frustrating to spend time on questions than
  can easily be solved by more or less elaborate techniques. As we now
  provide a PVS library for interval arithmetic and for Taylor models,
  one should be able  to answer quickly  to most of the easy questions
  about round-off, truncation and modeling errors. Concentrating only
  on  intricate  questions  is rewarding   from  the academia  and  ensures
  financial support from the industry.
\end{itemize}




In the future, we will  implement more operations on
Taylor models like square root, sine, cosine, and
arctangent. We will also create 
 PVS strategies to hide more and more details of Taylor models to users.
Our main goal remains to help provide {\em invisible formal methods}.

\acknowledgements

The authors wish to express all their gratitude to Cesar Muñoz from
the National Institute of Aerospace in Hampton, Virginia, for his tutoring and help
in the many manipulations around PVS. The authors would also like to thank NASA
Langley Research Center for its free PVS Class held on May 24-27, 2005.

\bibliographystyle{named}
\bibliography{alias,perso,groupe,saao,these,livre,arith}


         

    
\end{article}
\end{document}

%% file: finite_support.tex
\def\setsotherremovetwofn#1#2{{(#2 \setminus \{#1\})}}
\def\setsotheraddtwofn#1#2{{(#2 \cup \{#1\})}}
\def\setsotherdifferencetwofn#1#2{{(#1 \setminus #2)}}
\def\setsothercomplementonefn#1{{\overline{#1}}}
\def\setsotherintersectiontwofn#1#2{{(#1 \cap #2)}}
\def\setsotheruniontwofn#1#2{{(#1 \cup #2)}}
\def\setsotherstrictunderscoresubsetothertwofn#1#2{{(#1 \subset #2)}}
\def\setsothersubsetothertwofn#1#2{{(#1 \subseteq #2)}}
\def\setsothermembertwofn#1#2{{(#1 \in #2)}}
\def\opohtwofn#1#2{{#1\circ#2}}
\def\opdividetwofn#1#2{{#1/#2}}
\def\optimestwofn#1#2{{#1\times#2}}
\def\opdifferenceonefn#1{{-#1}}
\def\opdifferencetwofn#1#2{{#1-#2}}
\def\opplustwofn#1#2{{#1+#2}}
\begin{alltt}
\pvsid{finite\_support}: \pvskey{THEORY}
 \pvskey{BEGIN}

  \pvskey{IMPORTING} \pvsid{series}@\pvsid{series}, \pvsid{reals}@\pvsid{sqrt}, \pvsid{series}@\pvsid{power\_series}

  \(a\), \(b\), \(c\): \pvskey{VAR} \pvsid{sequence}\({\pvsbracketl}\)\pvsid{real}\({\pvsbracketr}\)\vspace*{\pvsdeclspacing}
  \(N\), \(M\), \(L\), \(n\), \(m\), \(l\), \(i\), \(j\): \pvskey{VAR} \pvsid{nat}\vspace*{\pvsdeclspacing}
  \(x\): \pvskey{VAR} \pvsid{real}\vspace*{\pvsdeclspacing}

  \pvsid{finite\_support}\pvsid{(}\(a\): \pvsid{sequence}\({\pvsbracketl}\)\pvsid{real}\({\pvsbracketr}\), \(N\): \pvsid{nat}\pvsid{)}: \pvsid{boolean} \pvskey{=}
      \(\forall\) \pvsid{(}\(n\): \pvsid{nat}\pvsid{)}: \(n\) \(>\) \(N\) \(\Rightarrow\) \(a\)\pvsid{(}\(n\)\pvsid{)} \(=\) \(0\)\vspace*{\pvsdeclspacing}

  \pvsid{cauchy}\pvsid{(}\(a\), \(b\): \pvsid{sequence}\({\pvsbracketl}\)\pvsid{real}\({\pvsbracketr}\)\pvsid{)}\pvsid{(}\(n\): \pvsid{nat}\pvsid{)}: \pvsid{real} \pvskey{=}
      \(\Sigma\)\pvsid{(}\(0\), \(n\),
          \(\lambda\) \pvsid{(}\(k\): \pvsid{nat}\pvsid{)}:
            \pvskey{IF} \(n\) \(\geq\) \(k\)
              \pvskey{THEN} \(\optimestwofn{a\pvsid{(}k\pvsid{)}}{b\pvsid{(}\opdifferencetwofn{n}{k}\pvsid{)}}\)
            \pvskey{ELSE} \(0\)
            \pvskey{ENDIF}\pvsid{)}\vspace*{\pvsdeclspacing}

  \pvsid{comp}\pvsid{(}\(a\), \(b\): \pvsid{sequence}\({\pvsbracketl}\)\pvsid{real}\({\pvsbracketr}\), \(d\): \pvsid{nat}\pvsid{)}: \pvskey{RECURSIVE} \pvsid{sequence}\({\pvsbracketl}\)\pvsid{real}\({\pvsbracketr}\) \pvskey{=}
    \pvskey{IF} \(d\) \(=\) \(0\)
      \pvskey{THEN} \pvsid{(}\(\lambda\) \(n\): \pvskey{IF} \(n\) \(=\) \(0\) \pvskey{THEN} \(a\)\pvsid{(}\(0\)\pvsid{)} \pvskey{ELSE} \(0\) \pvskey{ENDIF}\pvsid{)}
    \pvskey{ELSE} \pvskey{LET} \(c\) \pvskey{=} \pvsid{(}\(\lambda\) \(n\): \pvskey{IF} \(n\) \(=\) \(d\) \pvskey{THEN} \(0\) \pvskey{ELSE} \(a\)\pvsid{(}\(n\)\pvsid{)} \pvskey{ENDIF}\pvsid{)} \pvskey{IN}
           \(\opplustwofn{\optimestwofn{a\pvsid{(}d\pvsid{)}}{\pvsid{pow}\pvsid{(}b, d\pvsid{)}}}{\pvsid{comp}\pvsid{(}c, b, \opdifferencetwofn{d}{1}\pvsid{)}}\)
    \pvskey{ENDIF}
     \pvskey{MEASURE} \(d\)\vspace*{\pvsdeclspacing}

  \pvsid{neg\_fs}: \pvskey{LEMMA}
    \pvsid{finite\_support}\pvsid{(}\(a\), \(N\)\pvsid{)} \(\Rightarrow\) \pvsid{finite\_support}\pvsid{(}\(\opdifferenceonefn{a}\), \(N\)\pvsid{)}\vspace*{\pvsdeclspacing}
  \pvsid{add\_fs}: \pvskey{LEMMA}
    \pvsid{finite\_support}\pvsid{(}\(a\), \(N\)\pvsid{)} \(\wedge\) \pvsid{finite\_support}\pvsid{(}\(b\), \(M\)\pvsid{)} \(\wedge\) \(L\) \(\geq\) \pvsid{max}\pvsid{(}\(N\), \(M\)\pvsid{)} \(\Rightarrow\)
     \pvsid{finite\_support}\pvsid{(}\(\opplustwofn{a}{b}\), \(L\)\pvsid{)}\vspace*{\pvsdeclspacing}
  \pvsid{scal\_fs}: \pvskey{LEMMA}
    \pvsid{finite\_support}\pvsid{(}\(a\), \(N\)\pvsid{)} \(\Rightarrow\) \pvsid{finite\_support}\pvsid{(}\(\optimestwofn{x}{a}\), \(N\)\pvsid{)}\vspace*{\pvsdeclspacing}
  \pvsid{finite\_support\_mult}: \pvskey{LEMMA}
    \pvsid{finite\_support}\pvsid{(}\(a\), \(N\)\pvsid{)} \(\wedge\) \pvsid{finite\_support}\pvsid{(}\(b\), \(M\)\pvsid{)} \(\Rightarrow\)
     \pvsid{finite\_support}\pvsid{(}\pvsid{cauchy}\pvsid{(}\(a\), \(b\)\pvsid{)}, \(\opplustwofn{N}{M}\)\pvsid{)}\vspace*{\pvsdeclspacing}
  \pvsid{finite\_support\_cauchy}: \pvskey{LEMMA}
    \pvsid{finite\_support}\pvsid{(}\(a\), \(N\)\pvsid{)} \(\wedge\) \pvsid{finite\_support}\pvsid{(}\(b\), \(M\)\pvsid{)} \(\Rightarrow\)
     \(\optimestwofn{\pvsid{series}\pvsid{(}a\pvsid{)}\pvsid{(}N\pvsid{)}}{\pvsid{series}\pvsid{(}b\pvsid{)}\pvsid{(}M\pvsid{)}}\) \(=\)
      \pvsid{series}\pvsid{(}\pvsid{cauchy}\pvsid{(}\(a\), \(b\)\pvsid{)}\pvsid{)}\pvsid{(}\(\opplustwofn{N}{M}\)\pvsid{)}\vspace*{\pvsdeclspacing}
  \pvsid{finite\_support\_comp}: \pvskey{LEMMA}
    \pvsid{finite\_support}\pvsid{(}\(a\), \(N\)\pvsid{)} \(\wedge\) \pvsid{finite\_support}\pvsid{(}\(b\), \(M\)\pvsid{)} \(\Rightarrow\)
     \pvsid{finite\_support}\pvsid{(}\pvsid{comp}\pvsid{(}\(a\), \(b\), \(N\)\pvsid{)}, \(\optimestwofn{N}{M}\)\pvsid{)}\vspace*{\pvsdeclspacing}

 \pvskey{END} \pvsid{finite\_support}\end{alltt}

%% file: polynomials_ext.tex
\def\setsotherremovetwofn#1#2{{(#2 \setminus \{#1\})}}
\def\setsotheraddtwofn#1#2{{(#2 \cup \{#1\})}}
\def\setsotherdifferencetwofn#1#2{{(#1 \setminus #2)}}
\def\setsothercomplementonefn#1{{\overline{#1}}}
\def\setsotherintersectiontwofn#1#2{{(#1 \cap #2)}}
\def\setsotheruniontwofn#1#2{{(#1 \cup #2)}}
\def\setsotherstrictunderscoresubsetothertwofn#1#2{{(#1 \subset #2)}}
\def\setsothersubsetothertwofn#1#2{{(#1 \subseteq #2)}}
\def\setsothermembertwofn#1#2{{(#1 \in #2)}}
\def\opohtwofn#1#2{{#1\circ#2}}
\def\opdividetwofn#1#2{{#1/#2}}
\def\optimestwofn#1#2{{#1\times#2}}
\def\opdifferenceonefn#1{{-#1}}
\def\opdifferencetwofn#1#2{{#1-#2}}
\def\opplustwofn#1#2{{#1+#2}}
\begin{alltt}
\pvsid{polynomials\_ext}: \pvskey{THEORY}
 \pvskey{BEGIN}

  \pvskey{IMPORTING} \pvsid{finite\_support}, \pvsid{trig\_fnd}@\pvsid{polynomial\_deriv}

  \(a\), \(b\), \(d\): \pvskey{VAR} \pvsid{sequence}\({\pvsbracketl}\)\pvsid{real}\({\pvsbracketr}\)\vspace*{\pvsdeclspacing}
  \(n\), \(N\), \(M\), \(L\): \pvskey{VAR} \pvsid{nat}\vspace*{\pvsdeclspacing}
  \(c\): \pvskey{VAR} \pvsid{real}\vspace*{\pvsdeclspacing}
  \(x\), \(y\): \pvskey{VAR} \pvsid{real}\vspace*{\pvsdeclspacing}

  \pvsid{fs\_powerseq}: \pvskey{LEMMA}
    \pvsid{finite\_support}\pvsid{(}\(a\), \(N\)\pvsid{)} \(\Rightarrow\) \pvsid{finite\_support}\pvsid{(}\pvsid{powerseq}\pvsid{(}\(a\), \(x\)\pvsid{)}, \(N\)\pvsid{)}\vspace*{\pvsdeclspacing}

  \pvsid{fs\_condition}: \pvskey{LEMMA}
    \pvsid{finite\_support}\pvsid{(}\(a\), \(N\)\pvsid{)} \(\Rightarrow\)
     \pvsid{(}\(\forall\) \pvsid{(}\(i\): \pvsid{posnat}\pvsid{)}: \(a\)\pvsid{(}\(\opplustwofn{N}{i}\)\pvsid{)} \(=\) \(0\)\pvsid{)}\vspace*{\pvsdeclspacing}

  \pvsid{scal\_polynomial1}: \pvskey{LEMMA}
    \(\optimestwofn{x}{\pvsid{polynomial}\pvsid{(}a, N\pvsid{)}}\) \(=\) \pvsid{polynomial}\pvsid{(}\(\optimestwofn{x}{a}\), \(N\)\pvsid{)}\vspace*{\pvsdeclspacing}

  \pvsid{powerseries\_polynomial}: \pvskey{LEMMA}
    \pvsid{polynomial}\pvsid{(}\(a\), \(n\)\pvsid{)}\pvsid{(}\(x\)\pvsid{)} \(=\) \pvsid{powerseries}\pvsid{(}\(a\)\pvsid{)}\pvsid{(}\(x\)\pvsid{)}\pvsid{(}\(n\)\pvsid{)}\vspace*{\pvsdeclspacing}

  \pvsid{polynomial\_zero}: \pvskey{LEMMA}
    \pvsid{polynomial}\pvsid{(}\pvsid{(}\(\lambda\) \pvsid{(}\(n\): \pvsid{nat}\pvsid{)}: \(0\)\pvsid{)}, \(N\)\pvsid{)}\pvsid{(}\(x\)\pvsid{)} \(=\) \(0\)\vspace*{\pvsdeclspacing}

  \pvsid{mul\_polynomial}: \pvskey{LEMMA}
    \pvsid{finite\_support}\pvsid{(}\(a\), \(N\)\pvsid{)} \(\wedge\) \pvsid{finite\_support}\pvsid{(}\(b\), \(M\)\pvsid{)} \(\Rightarrow\)
     \(\optimestwofn{\pvsid{polynomial}\pvsid{(}a, N\pvsid{)}\pvsid{(}x\pvsid{)}}{\pvsid{polynomial}\pvsid{(}b, M\pvsid{)}\pvsid{(}x\pvsid{)}}\) \(=\)
      \pvsid{polynomial}\pvsid{(}\pvsid{cauchy}\pvsid{(}\(a\), \(b\)\pvsid{)}, \(\opplustwofn{N}{M}\)\pvsid{)}\pvsid{(}\(x\)\pvsid{)}\vspace*{\pvsdeclspacing}

  \pvsid{pow\_polynomial}: \pvskey{LEMMA}
    \pvsid{finite\_support}\pvsid{(}\(a\), \(N\)\pvsid{)} \(\Rightarrow\)
     \pvsid{polynomial}\pvsid{(}\(a\), \(N\)\pvsid{)}\pvsid{(}\(x\)\pvsid{)} \({}^{\scriptscriptstyle\wedge}\) \(n\) \(=\)
      \pvsid{polynomial}\pvsid{(}\pvsid{pow}\pvsid{(}\(a\), \(n\)\pvsid{)}, \(\optimestwofn{n}{N}\)\pvsid{)}\pvsid{(}\(x\)\pvsid{)}\vspace*{\pvsdeclspacing}

  \pvsid{comp\_polynomial}: \pvskey{LEMMA}
    \pvsid{finite\_support}\pvsid{(}\(a\), \(N\)\pvsid{)} \(\wedge\) \pvsid{finite\_support}\pvsid{(}\(b\), \(M\)\pvsid{)} \(\Rightarrow\)
     \pvsid{polynomial}\pvsid{(}\(a\), \(N\)\pvsid{)}\pvsid{(}\pvsid{polynomial}\pvsid{(}\(b\), \(M\)\pvsid{)}\pvsid{(}\(x\)\pvsid{)}\pvsid{)} \(=\)
      \pvsid{polynomial}\pvsid{(}\pvsid{comp}\pvsid{(}\(a\), \(b\), \(N\)\pvsid{)}, \(\optimestwofn{N}{M}\)\pvsid{)}\pvsid{(}\(x\)\pvsid{)};\vspace*{\pvsdeclspacing}

  \pvsid{geom\_polynomial}: \pvskey{LEMMA}
    \(\optimestwofn{\pvsid{(}\opdifferencetwofn{1}{x}\pvsid{)}}{\Sigma\pvsid{(}0, N, \lambda \pvsid{(}i: \pvsid{nat}\pvsid{)}: x {}^{\scriptscriptstyle\wedge} i\pvsid{)}}\) \(=\)
     \(\opdifferencetwofn{1}{x {}^{\scriptscriptstyle\wedge} \pvsid{(}\opplustwofn{N}{1}\pvsid{)}}\)\vspace*{\pvsdeclspacing}

 \pvskey{END} \pvsid{polynomials\_ext}\end{alltt}

%% file: taylor_model.tex
\def\setsotherremovetwofn#1#2{{(#2 \setminus \{#1\})}}
\def\setsotheraddtwofn#1#2{{(#2 \cup \{#1\})}}
\def\setsotherdifferencetwofn#1#2{{(#1 \setminus #2)}}
\def\setsothercomplementonefn#1{{\overline{#1}}}
\def\setsotherintersectiontwofn#1#2{{(#1 \cap #2)}}
\def\setsotheruniontwofn#1#2{{(#1 \cup #2)}}
\def\setsotherstrictunderscoresubsetothertwofn#1#2{{(#1 \subset #2)}}
\def\setsothersubsetothertwofn#1#2{{(#1 \subseteq #2)}}
\def\setsothermembertwofn#1#2{{(#1 \in #2)}}
\def\opohtwofn#1#2{{#1\circ#2}}
\def\opdividetwofn#1#2{{#1/#2}}
\def\optimestwofn#1#2{{#1\times#2}}
\def\opdifferenceonefn#1{{-#1}}
\def\opdifferencetwofn#1#2{{#1-#2}}
\def\opplustwofn#1#2{{#1+#2}}
\begin{alltt}
\pvsid{taylor\_model}\({\pvsbracketl}\)\(N\): \pvsid{nat}, \pvsid{(}\pvskey{IMPORTING} \pvsid{interval}@\pvsid{interval}\pvsid{)} \pvsid{domInterval}: \pvsid{Interval}\({\pvsbracketr}\): \pvskey{THEORY}
 \pvskey{BEGIN}

  \pvsid{tm}: \pvskey{TYPE} = \({\pvsrectypel}\)\(P\): \pvsid{fs\_type}, \(I\): \pvsid{Interval}\({\pvsrectyper}\)\vspace*{\pvsdeclspacing}

  \pvsid{tm\_equal}: \pvskey{AXIOM}
    \(t\) \(=\) \(u\) \(\equiv\)
     \pvsid{polynomial}\pvsid{(}\(t\)`\(P\), \(N\)\pvsid{)} \(=\) \pvsid{polynomial}\pvsid{(}\(u\)`\(P\), \(N\)\pvsid{)} \(\land\) \(t\)`\(I\) \(=\) \(u\)`\(I\);\vspace*{\pvsdeclspacing}

  \(\opplustwofn{t}{u: \pvsid{tm}}\): \pvsid{tm} \pvskey{=}  \pvsid{(#}\(P\) \pvskey{:=} \(\opplustwofn{t`P}{u`P}\), \(I\) \pvskey{:=} \(\opplustwofn{t`I}{u`I}\)\pvsid{#)};\vspace*{\pvsdeclspacing}
  \(\opdifferenceonefn{t}\): \pvsid{tm} \pvskey{=} \pvsid{(#}\(P\) \pvskey{:=} \(\opdifferenceonefn{t`P}\), \(I\) \pvskey{:=} \(\opdifferenceonefn{t`I}\)\pvsid{#)};\vspace*{\pvsdeclspacing}
  \(\optimestwofn{c}{t}\): \pvsid{tm} \pvskey{=} \pvsid{(#}\(P\) \pvskey{:=} \(\optimestwofn{c}{t`P}\), \(I\) \pvskey{:=} \(\optimestwofn{{\pvsbrackvbarl}c{\pvsbrackvbarr}}{t`I}\)\pvsid{#)}\vspace*{\pvsdeclspacing}
  \(\optimestwofn{t}{u}\): \pvsid{tm} \pvskey{=} \pvsid{(#}\(P\) \pvskey{:=} \pvsid{trunc}\pvsid{(}\pvsid{cauchy}\pvsid{(}\(t\)`\(P\), \(u\)`\(P\)\pvsid{)}, \(N\)\pvsid{)}, \(I\) \pvskey{:=} ... \pvsid{#)}\vspace*{\pvsdeclspacing}
  \pvsid{inv}\pvsid{(}\(t\): \{\(t\): \pvsid{tm} | same condition as below tm_inv_sharp \}\pvsid{)}:
        \pvsid{tm} \pvskey{=} \pvsid{(#}\(P\) \pvskey{:=} ... , \(I\) \pvskey{:=} ... \pvsid{#)}\vspace*{\pvsdeclspacing}

  \pvsid{containment}\pvsid{(}\(f\): \({\pvsbracketl}\)\pvsid{domIntervalType} \(\to\) \pvsid{real}\({\pvsbracketr}\), \(t\): \pvsid{tm}\pvsid{)}: \pvsid{bool} \pvskey{=}
      \(\forall\) \pvsid{xu}: \(\pvsid{(}\opdifferencetwofn{f\pvsid{(}\pvsid{xu}\pvsid{)}}{\pvsid{polynomial}\pvsid{(}t`P, N\pvsid{)}\pvsid{(}\pvsid{xu}\pvsid{)}}\pvsid{)}\) \pvsid{##} \(t\)`\(I\)\vspace*{\pvsdeclspacing}

  \pvsid{tm\_add\_sharp}: \pvskey{LEMMA}
    \pvsid{containment}\pvsid{(}\(f\), \(t\)\pvsid{)} \(\land\) \pvsid{containment}\pvsid{(}\(g\), \(u\)\pvsid{)} \(\Rightarrow\) \pvsid{containment}\pvsid{(}\(\opplustwofn{f}{g}\), \(\opplustwofn{t}{u}\)\pvsid{)}\vspace*{\pvsdeclspacing}
  \pvsid{tm\_scal\_sharp}: \pvskey{LEMMA}
    \pvsid{containment}\pvsid{(}\(f\), \(t\)\pvsid{)} \(\Rightarrow\) \pvsid{containment}\pvsid{(}\(\optimestwofn{x}{f}\), \(\optimestwofn{x}{t}\)\pvsid{)}\vspace*{\pvsdeclspacing}
  \pvsid{tm\_neg\_sharp}: \pvskey{LEMMA}
    \pvsid{containment}\pvsid{(}\(f\), \(t\)\pvsid{)} \(\Rightarrow\) \pvsid{containment}\pvsid{(}\(\opdifferenceonefn{f}\), \(\opdifferenceonefn{t}\)\pvsid{)}\vspace*{\pvsdeclspacing}
  \pvsid{tm\_mult\_sharp}: \pvskey{LEMMA}
    \pvsid{containment}\pvsid{(}\(f\), \(t\)\pvsid{)} \(\land\) \pvsid{containment}\pvsid{(}\(g\), \(u\)\pvsid{)} \(\Rightarrow\) \pvsid{containment}\pvsid{(}\(\optimestwofn{f}{g}\), \(\optimestwofn{t}{u}\)\pvsid{)}\vspace*{\pvsdeclspacing}
  \pvsid{tm\_inv\_sharp}: \pvskey{LEMMA}
    \(\forall\) \pvsid{(}\(f\): \({\pvsbracketl}\)\pvsid{domIntervalType} \(\to\) \pvsid{nzreal}\({\pvsbracketr}\),
         \(t\): \{\(t\): \pvsid{tm} |
                     \(t\)`\(P\)\pvsid{(}\(0\)\pvsid{)} \(\neq\) \(0\) \(\land\)
                      \(\pvsid{(}\opdividetwofn{t`I}{\pvsid{intervalFromRealSeq}\pvsid{(}t`P, N\pvsid{)}}\pvsid{)}\)`\pvsid{lb} \(\neq\) \(0\) \(\land\)
                       \(\pvsid{(}\opdividetwofn{t`I}{\pvsid{intervalFromRealSeq}\pvsid{(}t`P, N\pvsid{)}}\pvsid{)}\)`\pvsid{ub} \(\neq\) \(0\) \(\land\)
                        \(\pvsid{(}\opdividetwofn{t`I}{\pvsid{intervalFromRealSeq}\pvsid{(}t`P, N\pvsid{)}}\pvsid{)}\) \(>\) \(\opdifferenceonefn{1}\)\}\pvsid{)}:
      \pvsid{(}\(\forall\) \pvsid{xu}:
          \pvsid{polynomial}\pvsid{(}\(t\)`\(P\), \(N\)\pvsid{)}\pvsid{(}\pvsid{xu}\pvsid{)} \(\neq\) \(0\) \(\land\)
           \(\opdividetwofn{\pvsid{(}\opdifferencetwofn{f\pvsid{(}\pvsid{xu}\pvsid{)}}{\pvsid{polynomial}\pvsid{(}t`P, N\pvsid{)}\pvsid{(}\pvsid{xu}\pvsid{)}}\pvsid{)}}{\pvsid{polynomial}\pvsid{(}t`P, N\pvsid{)}\pvsid{(}\pvsid{xu}\pvsid{)}}\)
            \(\neq\) \(1\)
            \(\land\)
            \pvsid{polynomial}\pvsid{(}\(\lambda\) \pvsid{(}\(i\): \pvsid{nat}\pvsid{)}:
                          \pvskey{IF} \(i\) \(=\) \(0\) \pvskey{THEN} \(0\) \pvskey{ELSE} \(\opdividetwofn{\opdifferenceonefn{t`P\pvsid{(}i\pvsid{)}}}{t`P\pvsid{(}0\pvsid{)}}\) \pvskey{ENDIF},
                        \(N\)\pvsid{)}
                      \pvsid{(}\pvsid{xu}\pvsid{)}
             \(\neq\) \(1\)\pvsid{)}
       \(\land\) \pvsid{Zeroless?}\pvsid{(}\({\pvsbrackvbarl}\)\(t\)`\(P\)\pvsid{(}\(0\)\pvsid{)}\({\pvsbrackvbarr}\)\pvsid{)} \(\land\) \pvsid{Zeroless?}\pvsid{(} ... \pvsid{)}
         \(\land\) \pvsid{Zeroless?}\pvsid{(}\pvsid{intervalFromRealSeq}\pvsid{(}\(t\)`\(P\), \(N\)\pvsid{)}\pvsid{)} \(\land\) \pvsid{containment}\pvsid{(}\(f\), \(t\)\pvsid{)}
       \(\Rightarrow\) \pvsid{containment}\pvsid{(}\(\opdividetwofn{1}{f}\), \pvsid{inv}\pvsid{(}\(t\)\pvsid{)}\pvsid{)}\vspace*{\pvsdeclspacing}

 \pvskey{END} \pvsid{taylor\_model}\end{alltt}

%% file: examples.tex
\def\setsotherremovetwofn#1#2{{(#2 \setminus \{#1\})}}
\def\setsotheraddtwofn#1#2{{(#2 \cup \{#1\})}}
\def\setsotherdifferencetwofn#1#2{{(#1 \setminus #2)}}
\def\setsothercomplementonefn#1{{\overline{#1}}}
\def\setsotherintersectiontwofn#1#2{{(#1 \cap #2)}}
\def\setsotheruniontwofn#1#2{{(#1 \cup #2)}}
\def\setsotherstrictunderscoresubsetothertwofn#1#2{{(#1 \subset #2)}}
\def\setsothersubsetothertwofn#1#2{{(#1 \subseteq #2)}}
\def\setsothermembertwofn#1#2{{(#1 \in #2)}}
\def\opohtwofn#1#2{{#1\circ#2}}
\def\opdividetwofn#1#2{{#1/#2}}
\def\optimestwofn#1#2{{#1\times#2}}
\def\opdifferenceonefn#1{{-#1}}
\def\opdifferencetwofn#1#2{{#1-#2}}
\def\opplustwofn#1#2{{#1+#2}}
\begin{alltt}
\pvsid{example}: \pvskey{THEORY}
 \pvskey{BEGIN}

  \pvskey{IMPORTING} \pvsid{tm\_exp}\({\pvsbracketl}\)\(5\), \(5\), \pvsid{(#}\pvsid{lb} \pvskey{:=} \(\opdifferenceonefn{1}\), \pvsid{ub} \pvskey{:=} \(1\)\pvsid{#)}\({\pvsbracketr}\)

  \pvsid{ch}\pvsid{(}\(x\): \pvsid{tm}\pvsid{)}: \pvsid{tm} \pvskey{=}
      \(\optimestwofn{\pvsid{(}\opdividetwofn{1}{2}\pvsid{)}}{\pvsid{(}\opplustwofn{\pvsid{exp}\pvsid{(}x\pvsid{)}}{\pvsid{exp}\pvsid{(}\opdifferenceonefn{x}\pvsid{)}}\pvsid{)}}\)\vspace*{\pvsdeclspacing}

  \pvsid{sh}\pvsid{(}\(x\): \pvsid{tm}\pvsid{)}: \pvsid{tm} \pvskey{=}
      \(\optimestwofn{\pvsid{(}\opdividetwofn{1}{2}\pvsid{)}}{\pvsid{(}\opplustwofn{\pvsid{exp}\pvsid{(}x\pvsid{)}}{\opdifferenceonefn{\pvsid{exp}\pvsid{(}\opdifferenceonefn{x}\pvsid{)}}}\pvsid{)}}\)\vspace*{\pvsdeclspacing}

  \pvsid{seq\_px}: \pvsid{fs\_type} \pvskey{=}
      \(\lambda\) \pvsid{(}\(n\): \pvsid{nat}\pvsid{)}: \pvskey{IF} \(n\) \(=\) \(1\) \pvskey{THEN} \(\opdividetwofn{1}{1000}\) \pvskey{ELSE} \(0\) \pvskey{ENDIF}\vspace*{\pvsdeclspacing}

  \pvsid{tm\_x}: \pvsid{tm} \pvskey{=} \pvsid{(#}\(P\) \pvskey{:=} \pvsid{seq\_px}, \(I\) \pvskey{:=} \({\pvsbrackvbarl}\)\(0\)\({\pvsbrackvbarr}\)\pvsid{#)}\vspace*{\pvsdeclspacing}

  \pvsid{example1}: \pvsid{tm} \pvskey{=} \(\optimestwofn{\pvsid{ch}\pvsid{(}\optimestwofn{2}{\pvsid{tm\_x}}\pvsid{)}}{\pvsid{sh}\pvsid{(}\optimestwofn{3}{\pvsid{tm\_x}}\pvsid{)}}\)\vspace*{\pvsdeclspacing}

 \pvskey{END} \pvsid{example}\end{alltt}

%% file: examplesTrace.tex

<PVSio> example1`P(0);

==> 

0

<PVSio> example1`P(1);

==> 

3/1000

<PVSio> example1`P(2);

==> 

0

<PVSio> example1`P(3);

==> 

21/2000000000

<PVSio> example1`P(4);

==> 

0

<PVSio> example1`P(5);

==> 

521/40000000000000000

<PVSio> example1`I;

==> 

(\# lb
     := -1996666003792920908077809559596469417049924988435

67542489125827927772468257695416279793105352103584647/38763

49604747870233132233643700469577302245603256513727240130672

32422339563866364336668581220000000000000000000000000000,

   ub
     := 1996666003792920908077809559596469417049924988435

67542489125827927772468257695416279793105352103584647/38763

49604747870233132233643700469577302245603256513727240130672

32422339563866364336668581220000000000000000000000000000 \#)

%% file: taylor.bbl
\begin{thebibliography}{}

\bibitem[\protect\citeauthoryear{Aczel}{1996}]{Acz96}
Amir~D. Aczel.
\newblock {\em Fermat's last theorem: unlocking the secret of an ancient
  mathematical problem}.
\newblock Four Walls Eight Windows, 1996.

\bibitem[\protect\citeauthoryear{Bertot and Casteran}{2004}]{BerCas04}
Yves Bertot and Pierre Casteran.
\newblock {\em Interactive Theorem Proving and Program Development}.
\newblock Springer-Verlag, 2004.

\bibitem[\protect\citeauthoryear{Daumas \bgroup \em et al.\egroup
  }{2005}]{DauMelMun05}
Marc Daumas, Guillaume Melquiond, and César Muñoz.
\newblock Guaranteed proofs using interval arithmetic.
\newblock In Paolo Montuschi and Eric Schwarz, editors, {\em Proceedings of the
  17th Symposium on Computer Arithmetic}, Cape Cod, Massachusetts, 2005.

\bibitem[\protect\citeauthoryear{Gage and McCormick}{2004}]{GagMcC04}
Debbie Gage and John McCormick.
\newblock We did nothing wrong.
\newblock {\em Baseline}, 1(28):32--58, 2004.

\bibitem[\protect\citeauthoryear{{Information Management and Technology
  Division}}{1992}]{USA.92}
{Information Management and Technology Division}.
\newblock Patriot missile defense: software problem led to system failure at
  {D}hahran, {S}audi {A}rabia.
\newblock Report B-247094, United States General Accounting Office, 1992.

\bibitem[\protect\citeauthoryear{Jaulin \bgroup \em et al.\egroup
  }{2001}]{JauKieDidWal01}
Luc Jaulin, Michel Kieffer, Olivier Didrit, and Eric Walter.
\newblock {\em Applied interval analysis}.
\newblock Springer, 2001.

\bibitem[\protect\citeauthoryear{Lions and others}{1996}]{Lio.96}
Jacques-Louis Lions et~al.
\newblock Ariane 5 flight 501 failure report by the inquiry board.
\newblock Technical report, European Space Agency, Paris, France, 1996.

\bibitem[\protect\citeauthoryear{Makino and Berz}{2003}]{MakBer03}
Kyoko Makino and Martin Berz.
\newblock Taylor models and other validated functional inclusion methods.
\newblock {\em International Journal of Pure and Applied Mathematics},
  4(4):379--456, 2003.

\bibitem[\protect\citeauthoryear{Moore}{1966}]{Moo66}
Ramon~E. Moore.
\newblock {\em Interval analysis}.
\newblock Prentice Hall, 1966.

\bibitem[\protect\citeauthoryear{Neumaier}{1990}]{Neu90}
Arnold Neumaier.
\newblock {\em Interval methods for systems of equations}.
\newblock Cambridge University Press, 1990.

\bibitem[\protect\citeauthoryear{Owre \bgroup \em et al.\egroup
  }{1992}]{OwrRusSha92}
Sam Owre, John~M. Rushby, and Natarajan Shankar.
\newblock {PVS}: a prototype verification system.
\newblock In Deepak Kapur, editor, {\em 11th International Conference on
  Automated Deduction}, pages 748--752, Saratoga, New-York, 1992.
  Springer-Verlag.

\bibitem[\protect\citeauthoryear{Owre \bgroup \em et al.\egroup
  }{2001a}]{OwrShaRusStr01b}
Sam Owre, Natarajan Shankar, John~M. Rushby, and David W.~J. Stringer-Calvert.
\newblock {\em PVS Language Reference}.
\newblock SRI International, 2001.
\newblock Version 2.4.

\bibitem[\protect\citeauthoryear{Owre \bgroup \em et al.\egroup
  }{2001b}]{OwrShaRusStr01a}
Sam Owre, Natarajan Shankar, John~M. Rushby, and David W.~J. Stringer-Calvert.
\newblock {\em PVS System Guide}.
\newblock SRI International, 2001.
\newblock Version 2.4.

\bibitem[\protect\citeauthoryear{Ross}{2005}]{Ros05}
Philip~E. Ross.
\newblock The exterminators.
\newblock {\em IEEE Spectrum}, 42(9):36--41, 2005.

\bibitem[\protect\citeauthoryear{Rushby and von Henke}{1991}]{RusHen91}
John Rushby and Friedrich von Henke.
\newblock Formal verification of algorithms for critical systems.
\newblock In {\em Proceedings of the Conference on Software for Critical
  Systems}, pages 1--15, New Orleans, Louisiana, 1991.

\bibitem[\protect\citeauthoryear{Tiwari \bgroup \em et al.\egroup
  }{2003}]{TiwShaRus03}
Ashish Tiwari, Natarajan Shankar, and John Rushby.
\newblock Invisible formal methods for embedded control systems.
\newblock {\em Proceedings of the IEEE}, 91(1):29--39, 2003.

\end{thebibliography}
